\documentclass[twocolumn]{openjournal}

\usepackage{lipsum}
\usepackage{amsmath}

\usepackage{xcolor}
\usepackage{textgreek}
\usepackage[utf8]{inputenc}
\usepackage[english]{babel}

\usepackage{hyperref}
\hypersetup{
    unicode, 
    colorlinks=true,
    linkcolor=linkcolor,
    citecolor=linkcolor,
    filecolor=linkcolor,
    urlcolor=linkcolor,
}
\usepackage{color,colortbl}
\definecolor{linkcolor}{rgb}{0.0,0.3,0.5}
\usepackage{tensind}
\tensordelimiter{?}
\DeclareGraphicsExtensions{.bmp,.png,.jpg,.pdf}
\usepackage{verbatim}
\usepackage[normalem]{ulem}
\usepackage{orcidlink}
\usepackage{soul}

\makeatletter
\def\@hex@@Hex#1%
 {\if a#1A\else \if b#1B\else \if c#1C\else \if d#1D\else
  \if e#1E\else \if f#1F\else #1\fi\fi\fi\fi\fi\fi \@hex@Hex}
\makeatother
\definecolor{afcolor}{HTML}{b3443c}

\graphicspath{ {./figs/} }

\def\be{\begin{equation}}
\def\ee{\end{equation}}

\def\gsim{\lower.5ex\hbox{\gtsima}} 
\def\lsim{\lower.5ex\hbox{\ltsima}} 
\def\gtsima{$\; \buildrel > \over \sim \;$} 
\def\ltsima{$\; \buildrel < \over \sim \;$} \def\gsim{\lower.5ex\hbox{\gtsima}} 
\def\lsim{\lower.5ex\hbox{\ltsima}} 
\def\simgt{\lower.5ex\hbox{\gtsima}} 
\def\simlt{\lower.5ex\hbox{\ltsima}}

\def\S*{$\Sigma_{\rm SFR}$}

\def\kms{{\rm km\,s}^{-1}}

\definecolor{apcolor}{HTML}{b3003b}
\definecolor{afcolor}{HTML}{800080}
\definecolor{lvcolor}{HTML}{DF7401}
\definecolor{mdcolor}{HTML}{01abdf} 
\definecolor{cbcolor}{HTML}{ff0000}
\definecolor{sccolor}{HTML}{cc5500} 
\definecolor{sgcolor}{HTML}{00cc7a}

\begin{document}
\title{\bf On the clumpy nature of super-early galaxies}
\author{Andrea Ferrara\orcidlink{0000-0002-9400-7312}}
\affiliation{Scuola Normale Superiore,  Piazza dei Cavalieri 7, Pisa I-56126, Italy}
\affiliation{Centro B. Segre, Accademia Nazionale dei Lincei, Via della Lungara, 10, Roma I-00165, Italy}

\email{andrea.ferrara@sns.it}
\author{Barnali Das\orcidlink{0009-0004-2041-1023}}
\affiliation{Scuola Normale Superiore,  Piazza dei Cavalieri 7, Pisa I-56126, Italy}
\author{Mahsa Kohandel\orcidlink{0000-0003-1041-7865}}
\affiliation{INAF/OAS Bologna, Via Piero Gobetti 101 / Via Gobetti 93/3, 40129 Bologna, Italy}
\affiliation{Scuola Normale Superiore,  Piazza dei Cavalieri 7, Pisa I-56126, Italy}
\author{Andrea Pallottini\orcidlink{0000-0002-7129-5761}}
\affiliation{Dipartimento di Fisica {`Enrico Fermi'}, Universit\`{a} di Pisa, Largo Bruno Pontecorvo 3, Pisa I-56127, Italy}
\author{Evangelia Ntormousi\orcidlink{0000-0002-4324-0034}}
\affiliation{Scuola Normale Superiore,  Piazza dei Cavalieri 7, Pisa I-56126, Italy}

\begin{abstract}

The {\it James Webb Space Telescope} (JWST) has revealed that galaxies
during the Epoch of Reionization are composed of compact stellar clumps
spanning more than two orders of magnitude in mass and size. We present
an analytical framework that connects the global properties of
high-redshift galactic disks to the formation,
dynamical evolution and visibility of these systems. Starting from the classical Toomre instability, we derive analytical mass--size and
surface density--size relations together with, for the first time, the
intrinsic clump mass spectrum obtained by integrating over the entire
unstable branch of the dispersion relation. The observed clumps are
naturally reproduced for turbulent velocity dispersions
$\sigma\simeq10-80\,{\rm km\,s^{-1}}$, while their location along the
mass--size relation is determined by the compactness of the parent disk,
$\mathcal{C}$. The intrinsic spectrum extends over nearly
two decades in mass,
$5.5\lesssim\log(M_{\rm cl}/M_\odot)\lesssim7.3$, it is transformed by
gas dynamical friction, which suppresses the high-mass tail
($t_{\rm df}\propto M_{\rm cl}^{-1}$), and  { is consistent with} the observed clump
mass-function slope,
$dN/dM\propto M^{-1.89}$.
{
Massive Toomre fragments are resistant to all the stellar
feedback mechanisms considered, which generally become effective
only after $\gtrsim 95\%$ of the initial gas has been converted
into stars. This favors high star-formation efficiencies,
allowing the resulting stellar clumps to largely preserve
their initial Toomre masses.} The results provide the first self-consistent analytical framework connecting the formation, statistical properties and visibility of stellar clumps in super-early galaxies.
\end{abstract}

\maketitle

\section{Introduction}\label{sec:intro}
The unprecedented angular resolution and sensitivity of the {\it James Webb Space Telescope} (JWST) have revealed that galaxies in the first billion years of cosmic history are far from smooth stellar systems. 
Instead, many galaxies at $z\gtrsim6$ are resolved into multiple compact star-forming condensations with characteristic sizes of only a few tens to a few hundreds of parsecs. 

Gravitational lensing has been instrumental in reaching these physical scales, allowing individual stellar clumps to be resolved in intrinsically faint galaxies that would otherwise remain unresolved. 
Recent observations have identified galaxies at $z\simeq6-12$ whose stellar light is dominated by a handful of compact knots, suggesting that highly clustered star formation is a common mode of galaxy assembly in the early Universe \citep{Vanzella23,Adamo24b,Mowla24,Bradac25,Nakane25,Fujimoto25, Abdurrouf25, Messa26, Zhu26, Vanzella26}. These observations indicate that the clumpy morphology is not restricted to the most luminous systems but extends to galaxies with stellar masses $<10^8\, M_\odot$ \citep[][]{Adamo24b, Nakane25}, where the majority of cosmic star formation is expected to occur during the Epoch of Reionization \citep[EoR,][]{Atek25}.

Rather than representing merely morphological irregularities, these compact star-forming regions may constitute the fundamental building blocks through
which the earliest galaxies assemble their stellar mass. In this picture,
galaxy evolution proceeds through the formation, interaction, migration, and
eventual dissolution of individual clumps, whose collective evolution shapes
the global properties of the host galaxy.

The existence of massive star-forming clumps has important implications for virtually every aspect of early galaxy evolution. 
Clumps dominate the instantaneous star formation rate \citep[][]{Topping24, Morishita24, Chen26}, rapidly enrich the interstellar medium with metals and dust, and inject momentum and energy through stellar radiation, winds, and supernova explosions . 
Their subsequent dynamical evolution may redistribute angular momentum. 
In particular, inward migration driven by gravitational torques and dynamical
friction may transport both gas and stars toward the galactic centre,
contributing to the build-up of compact bulges or nuclear star clusters \citep[][]{Noguchi99} while
simultaneously redistributing angular momentum throughout the disk \citep[][]{Benton24, Hodge25}. Such mass
transport may also alter the disk surface density itself, thereby modifying
its gravitational stability and regulating subsequent episodes of clump
formation.
Whether these clumps survive as bound stellar systems \citep[][]{Krumholz10b}, dissolve into the diffuse stellar component \citep[][]{Meng22}, or migrate toward the galactic centre \citep[][]{Dekel21} therefore affects both the structural evolution of galaxies and their role during cosmic reionization. 

The physical origin of these clumps remains actively debated (for a discussion, see \citealt{Meng20,Zanella21}). One class of models attributes them to mergers and tidal interactions, whose frequency is expected to be high in the rapidly assembling halos characteristic of the early Universe \citep[][]{Puskas25,Duan26}. Cosmological simulations indeed show that major mergers can produce short-lived, intensely star-forming condensations that reproduce many observed morphologies \citep[e.g.][]{Nakazato24, Wang26}. 

An alternative picture invokes in-situ fragmentation of gas-rich, turbulent disks through gravitational instability. The conditions for early disk formation have been studied by \citet[][]{Kohandel25, Semenov26}; their existence has been shown by recent works \citep[e.g.][]{Rowland24, Carniani25, Cescon26}. In this scenario, first proposed for high-redshift galaxies well before the advent of JWST \citep[][]{Romeo92, Romeo11, Griv12, Romeo13, Romeo14, Leung20}, disks become unstable when the Toomre parameter, $Q$, approaches unity, leading to the growth of self-gravitating condensations whose characteristic scale is determined by the competition between self-gravity, pressure, and rotation \citep{Fisher17a, Fisher17b}. 

{The two formation channels need not be mutually exclusive. Indeed, Puskás et al. (in prep.), using JADES data, find that while major mergers can plausibly account for much of the observed clumpy fraction in low-mass, high-$z$ galaxies, an additional in-situ channel is required mainly at $M_\star\gtrsim10^9\,M_\odot$. Interactions and minor mergers can also perturb an already gas-rich, marginally unstable disk, locally enhancing its surface density and triggering subsequent in-situ fragmentation. Distinguishing such interaction-triggered Toomre clumps from accreting satellites may require resolved kinematic and chemical information: genuine disk fragments should approximately share the rotation and enrichment of the surrounding disk, whereas ex-situ systems may exhibit distinct orbital or stellar-population properties. We do not attempt such a classification here.}

However, observations do not reveal a single characteristic clump mass but
rather a population spanning $ > 2$ dex in mass and
size \citep[e.g.][]{Adamo24b, Nakane25}. Consequently, the relevant theoretical prediction is not only the mass
associated with the fastest-growing unstable mode, but the entire clump mass
function. Such a distribution contains substantially more information than a
single characteristic scale and provides a direct connection between the
physics of gravitational instability and the statistical properties of the
observed clump population.

Theoretical studies of giant clumps long predate the observational evidence
provided by JWST. The modern picture originated with the work of
\citet[][see also \citealt{Krumholz10, Dekel13}]{Dekel09}, who proposed that gas-rich, rapidly accreting disks at
high redshift naturally evolve toward marginal gravitational stability
($Q\simeq1$) and fragment through violent disk instability (VDI, \citealt{Mandelker17, Ceverino23}), producing
massive star-forming clumps whose characteristic masses are determined by the
global disk properties \citep[][]{Immeli04, Elmegreen08, Genzel08, Genzel11}. Shortly thereafter, numerical simulations confirmed
that such unstable disks indeed fragment into giant clumps which subsequently
interact, exchange angular momentum, and migrate toward the galactic centre,
where they contribute to bulge growth
\citep{Bournaud07,Ceverino10}.

Over the past decade, increasingly sophisticated cosmological simulations
have investigated the subsequent evolution of these clumps, including their
lifetimes, migration, stellar feedback, and survival
\citep[e.g.][]{Mandelker17,Dekel21,Dekel22, Fensch21, Mayer25, Horie26}. These studies
have established that clumps are not merely transient star-forming regions,
but may play a central role in driving the internal evolution of galaxies by
transporting mass and angular momentum and by regulating star formation
through feedback. 

At the same time, both simulations and observations have
suggested that the classical linear Toomre analysis may not provide a
complete description of the fragmentation process. Giant clumps have been
found to form even in regions where the local Toomre parameter exceeds unity,
pointing to the importance of nonlinear perturbations, compressive
turbulence, and cosmological gas accretion in triggering fragmentation
\citep{Inoue16,Krumholz16, Mandelker25}. Furthermore, observational limitations may
blend multiple smaller condensations into apparently massive clumps,
complicating direct comparisons between theory and observations
\citep{Meng20}.

Despite these important advances, most existing theoretical work has focused
either on predicting the characteristic mass associated with the
fastest-growing Toomre mode or on following the subsequent evolution of
individual clumps in numerical simulations. Comparatively little attention
has been devoted to deriving the {\it intrinsic mass spectrum} expected from
gravitational instability itself. Yet the mass function is a particularly informative observable now beginning to become accessible with JWST and carries considerably more information on the fragmentation physics.
Recently, analytic models have begun to address the statistical properties of
star-forming clumps in unstable disks. For example, 
\citet[][]{Orr24} focuses on the distribution of star-forming structures and star-formation properties, but a theoretical prediction directly connecting
the complete spectrum of unstable Toomre modes to the observable clump mass
distribution is still lacking.

The present work aims to fill this gap. Building upon the classical Toomre
instability framework, we derive the intrinsic clump mass function by
integrating over the entire unstable band of the dispersion relation, rather
than restricting the analysis to the fastest-growing mode alone. We then follow the subsequent dynamical evolution of this spectrum by
accounting for the preferential inward migration of massive clumps due
to gas dynamical friction. Finally, we investigate whether stellar
feedback can disperse the natal gas and dust surrounding the newly
formed systems, thereby allowing the resulting stellar clumps to become
observable.

The resulting framework connects the global properties of high-redshift
galactic disks to the statistical distribution of stellar clumps revealed by
JWST, providing a quantitative theoretical framework within which current and
future observations can be interpreted in the context of the concordance
$\Lambda$CDM cosmological model\footnote{We assume a flat Universe with the following cosmological parameters: $\Omega_m = 0.3075$, $\Omega_{\Lambda} = 1- \Omega_{\rm M}$, and $\Omega_{b} = 0.0486$,  $h=0.6774$, $\sigma_8=0.8159$, $n_s=0.9667$ where $\Omega_{m}$, $\Omega_{\Lambda}$, and $\Omega_{b}$ are the total matter, vacuum, and baryon densities, in units of the critical density; $h$ is the Hubble constant in units of $100\,\rm km\ s^{-1} Mpc^{-1}$, $\sigma_8$ is the late-time fluctuation amplitude parameter, and $n_s$ the spectral index. \citep{planck:2015}.}.

\section{Model}
\label{sec:formation}
\subsection{Disk model and Toomre stability}

We model the gas disk forming inside a dark matter halo of virial mass
$M_{\rm vir}$ and virial radius $R_{\rm vir}$. The total gas mass associated
with the halo is assumed to be
\begin{equation}
    M_g = f_b M_{\rm vir},
\end{equation}
where $f_b = \Omega_b/\Omega_m$ is the cosmic baryon fraction. We further assume that a fraction
$f_d$ of this gas settles into a rotationally supported disk, so that
\begin{equation}
    M_d = f_d f_b M_{\rm vir}.
\end{equation}

Following the standard angular-momentum-conserving disk model, we define
\begin{equation}
    m_d \equiv \frac{M_d}{M_{\rm vir}} = f_d f_b,
    \qquad
    j_d \equiv \frac{J_d}{J_{\rm vir}},
\end{equation}
where $J_d$ is the angular momentum of the disk and $J_{\rm vir}$ that of
the halo. If the gas that forms the disk conserves its specific angular
momentum during collapse, then
\begin{equation}
    \frac{J_d}{M_d} = \frac{J_{\rm vir}}{M_{\rm vir}},
\end{equation}
or equivalently
\begin{equation}
    \frac{j_d}{m_d}=1.
\end{equation}
The exponential disk scale radius is then
\begin{equation}
    R_d =
    \frac{1}{\sqrt{2}}
    \left(\frac{j_d}{m_d}\right)
    \lambda R_{\rm vir},
\end{equation}
where $\lambda$ is the halo spin parameter. In the fiducial case of specific
angular momentum conservation, this reduces to
\begin{equation}
    R_d = \frac{\lambda}{\sqrt{2}} R_{\rm vir}.
\end{equation}

We assume that the gas surface density follows an exponential radial profile,
\begin{equation}
    \Sigma_g(R) = \Sigma_0 \exp\left(-\frac{R}{R_d}\right),
\end{equation}
with normalization fixed by the total disk mass,
\begin{equation}
    M_d = 2\pi \Sigma_0 R_d^2.
\end{equation}
Therefore,
\begin{equation}
    \Sigma_0 = \frac{M_d}{2\pi R_d^2}
    = \frac{f_d f_b M_{\rm vir}}{2\pi R_d^2}.
\end{equation}

\begin{table}
\centering
\caption{Halo masses corresponding to $3\sigma$, $4\sigma$, and $5\sigma$ peaks of the linear density field for the adopted cosmology. The symbol (s) indicates that the disk is Toomre-stable. 
}
\renewcommand{\arraystretch}{1.4}
\begin{tabular}{clll}
\hline

$z$ &
$\log(M_{5\sigma}/M_\odot)$ &
$\log(M_{4\sigma}/M_\odot)$ &
$\log(M_{3\sigma}/M_\odot)$ \\

\hline\hline
15 &  9.45 (s) &  8.30 (s) &  6.54 (s) \\
14 &  9.75 (s)&  8.65 (s)&  6.97 (s)\\
13 & 10.05 (s)&  9.01 (s)&  7.41 (s)\\
12 & 10.37 (s)&  9.37 (s)&  7.85 (s)\\
11 & 10.69 &  9.74 (s)&  8.30 (s)\\
10 & 11.02 & 10.13 (s)&  8.77 (s)\\
9 & 11.37 & 10.53 (s)&  9.25 (s)\\
8 & 11.73 & 10.94 &  9.74 (s)\\
7 & 12.11 & 11.37 & 10.26 (s)\\
6 & 12.51 & 11.82 & 10.80 (s)\\
5 & 12.93 & 12.30 & 11.37 \\
\hline
\end{tabular}
\label{tab:sigma_peak_halo_masses}
\end{table}
The dark matter halo is assumed to follow a Navarro--Frenk--White
radial profile,
\begin{equation}
\rho_{\rm NFW}(r)
=
\frac{\rho_s}
{(r/r_s)(1+r/r_s)^2},
\end{equation}
where $r$ denotes the spherical halo radius and
\begin{equation}
r_s=\frac{R_{\rm vir}}{c_{\rm NFW}}
\end{equation}
is the halo scale radius.

The mass enclosed within a spherical radius $r$ is
\begin{equation}
M_{\rm NFW}(<r)
=
M_{\rm vir}
\frac{f(r/r_s)}{f(c_{\rm NFW})},
\end{equation}
where
\begin{equation}
f(x)=\ln(1+x)-\frac{x}{1+x}.
\end{equation}

To compute the gravitational potential experienced by the disk, we
evaluate the halo quantities in the disk midplane, where $r=R$.
The corresponding circular velocity is therefore
\begin{equation}
V_c^2(R)
=
\frac{G M_{\rm NFW}(<R)}{R},
\end{equation}
We compute the angular frequency as
\begin{equation}
    \Omega(R)=\frac{V_c(R)}{R}.
\end{equation}
For a spherical potential, using $\Omega^2=GM(<R)/R^3$, the epicyclic frequency can be written as
\begin{equation}
    \kappa^2(R)
    =
    \Omega(R)^2
    \left[
    1+\frac{{\rm d}\ln M_{\rm NFW}(<R)}{{\rm d}\ln R}
    \right].
\end{equation}
{For simplicity, we compute the epicyclic frequency from the dark-matter halo potential, neglecting the gravitational contribution of the baryonic disk. This approximation may become less accurate in the inner, baryon-dominated regions. For example, adopting a flat rotation curve with $V_c=V_{\rm vir}$ would increase $\kappa$ (and hence $Q$) by a factor of $\simeq1.6$ at $R_d$ for our representative galaxy, shifting the quantitative instability threshold without changing the underlying fragmentation mechanism.}
For the NFW profile,
\begin{equation}
    \frac{{\rm d}\ln M_{\rm NFW}}{{\rm d}\ln R}
    =
    \frac{x f'(x)}{f(x)},
    \qquad
    x \equiv \frac{R}{r_s}.
\end{equation}
Thus,
\begin{equation}
    \frac{{\rm d}\ln M_{\rm NFW}}{{\rm d}\ln R}
    =
    \frac{x^2}{(1+x)^2 f(x)}.
\end{equation}

The local Toomre parameter of the gas disk is then
\begin{equation}
    Q(R)
    =
    \frac{\sigma \kappa(R)}
    {\pi G \Sigma_g(R)},
\label{eq:ToomreQ}
\end{equation}
where $\sigma$ is the one-dimensional gas velocity dispersion, including
both thermal and turbulent support. More explicitly,
\begin{equation}
    \sigma^2
    =
    c_s^2+\sigma_{\rm turb}^2 .
\end{equation}
In high-redshift galaxies the turbulent component is expected to dominate \citep{Vallini18, Cescon26},
$\sigma_{\rm turb}\gg c_s$, so that $\sigma\simeq\sigma_{\rm turb}$.
In the fiducial calculations below we therefore regard $\sigma$ as the
effective turbulent velocity dispersion of the gas disk.

\begin{center}
    \begin{figure}[!t]
        \centering
    	\includegraphics[scale=0.34]{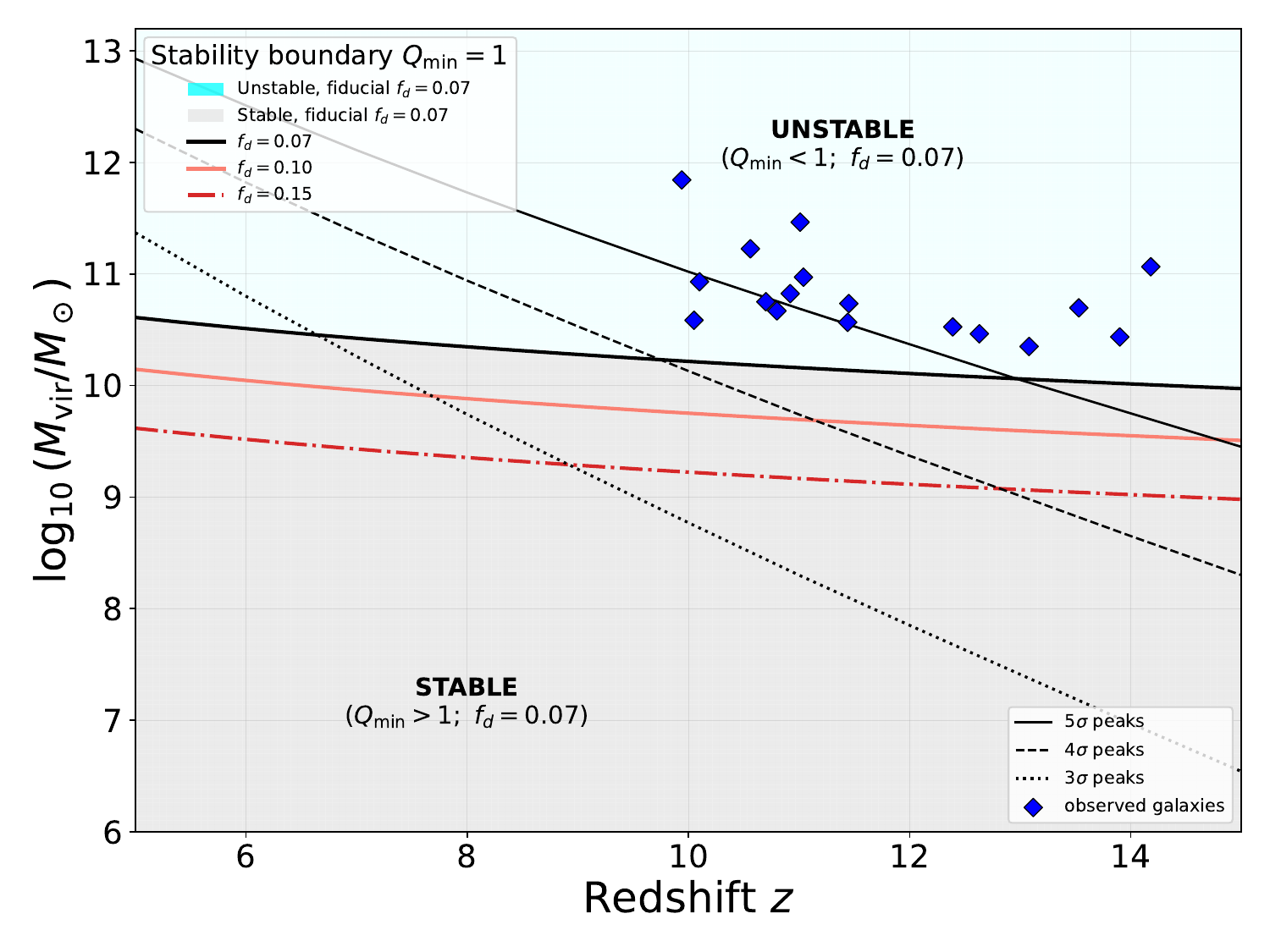}
    	\caption{Stability of gaseous disks in the halo mass--redshift plane. The azure and grey regions denote halos hosting Toomre-unstable ($Q_{\rm min} <1$) and Toomre-stable ($Q_{\rm min}>1$) disks, respectively, for the fiducial model ($f_d=0.07, \lambda=0.035, c_{\rm NFW}=5, \sigma \approx 10\,  \kms$); $Q_{\rm min}$ is the minimum value of the Toomre parameter computed over the radial range $0.05 R_d < R <5 R_d$. The thick black solid curve marks the transition $Q_{\rm min}=1$; orange solid and red dot-dashed curves correspond to $f_d=0.10, 0.15$, respectively. The thin black solid, dashed, and dotted curves show the halo masses corresponding to $5\sigma, 4\sigma$, and $3\sigma$ peaks of the primordial density field, respectively (Table~1). The blue points, showing the location of 17 super-early ($z>10$), spectroscopically confirmed galaxies are taken from \citet[][]{Ferrara26}. 
}
        \label{fig:stability}
    \end{figure}
\end{center}

This framework allows us to connect the global halo properties
$(M_{\rm vir}, R_{\rm vir}, c_{\rm NFW}, \lambda)$ to the local gravitational stability
of the resulting gas disk. Larger disk masses, smaller spin parameters, or
lower turbulence increase the disk surface density relative to pressure and
rotation, thereby reducing $Q$ and making the disk more susceptible to
axisymmetric fragmentation. Some of these dependences are quantitatively displayed 
in Fig. \ref{fig:stability}. 

{ For the fiducial model ($f_d=0.07$, $\sigma=10\,{\rm km\,s^{-1}}$),
the spectroscopically confirmed super-early galaxies shown in
Fig.~\ref{fig:stability} populate the Toomre-unstable region.
However, the instability threshold depends sensitively on the
turbulent velocity dispersion. Increasing $\sigma$ to
$30$ and $50\,{\rm km\,s^{-1}}$ reduces the number of unstable
systems from 17 to 1 and 0, respectively. Higher disk gas fractions
partly compensate for this stabilization: for $f_d=0.15$ and
$\sigma=30\,{\rm km\,s^{-1}}$, 11 of the 17 systems remain unstable.
Thus, fragmentation is favored in sufficiently gas-rich, compact
disks with moderate turbulent support, rather than being a universal
outcome for all super-early galaxies.}

This conclusion is also supported by high-resolution simulations.
For example, \citet{Kohandel20} found that the [CII]-emitting disk of
a simulated EoR galaxy has $\sigma_{\rm [CII]}\simeq23$--$38\,
{\rm km\,s^{-1}}$ and remains strongly gravitationally unstable,
with a characteristic Toomre parameter $Q\simeq0.2$.

We distinguish throughout between fiducial structural parameters of the
disk model and representative halo properties adopted for illustrative
calculations. We take $\lambda=0.035$ and $c_{\rm NFW}=5$ as fiducial structural
parameters, while the disk mass fraction $f_d$ and turbulent velocity
dispersion $\sigma$ are varied over the ranges motivated by Figs.~~\ref{fig:stability} and \ref{fig:Mcl_vs_Re}. In particular, the choice $f_d=0.07$ defines the fiducial stability
model shown in Fig.~\ref{fig:stability}, whereas the values
$M_{\rm vir}=7\times10^{10}\,M_\odot$ and $z=11$ adopted for the clump
mass-spectrum calculations in Figs.~\ref{fig:spectrum} and \ref{fig:spectrum_fd} should be regarded as a
representative super-early galaxy rather than as unique fiducial values. The sensitivity of the results to variations in these parameters is
discussed explicitly below where relevant.

\subsection{Fastest-growing Toomre mode and clump formation}
While $\Omega$ (or, equivalently, $\kappa$) characterizes the orbital response of the disk, the actual growth rate of the Toomre instability is determined by the full dispersion relation. For a thin gaseous disk, the local axisymmetric dispersion relation is
\begin{equation}
\omega^2(k,R)
=
\kappa^2(R)
-
2\pi G \Sigma_g(R), |k|
+
\sigma^2 k^2,
\end{equation}
where $k$ is the perturbation wavenumber. Although this relation has been generalized to multi-component (gas + stars) and/or finite-thickness disks \citep[e.g.,][]{Romeo11,Romeo13}, we retain the classical formulation. This choice is justified because the disks considered here are initially purely gaseous, and high-resolution simulations (Das et al., in prep.) show that they remain only moderately thick ($V_c/\sigma \approx 5$). Although a geometrically thin disk is not necessarily dynamically thin \citep[see, e.g.,][]{Grisdale17}, we neglect here this possible difference.

Instability occurs whenever $\omega^2(k,R) < 0$.
In this regime, perturbations grow exponentially with growth rate
\begin{equation}
    s(k,R)
    =
    \sqrt{-\omega^2(k,R)}.
\label{eq:growth_rate}
\end{equation}

At fixed radius, the fastest-growing mode is obtained by minimizing
$\omega^2$ with respect to $k$. This yields the fastest-growing wavenumber
\begin{equation}
    k_{\rm max}(R)
    =
    \frac{\pi G \Sigma_g(R)}{\sigma^2};
\end{equation}
the corresponding wavelength is therefore
\begin{equation}
    \lambda_{\rm max}(R)
    =
    \frac{2\pi}{k_{\rm max}}
    =
    \frac{2 \sigma^2}{G \Sigma_g(R)}.
\end{equation}

Substituting $k_{\rm max}$ into the dispersion relation gives the maximum local growth rate,
\begin{equation}
    s_{\rm max}(R)
    =
    \left[
    \left(
    \frac{\pi G \Sigma_g(R)}{\sigma}
    \right)^2
    -
    \kappa^2(R)
    \right]^{1/2}.
\end{equation}

Using the definition of the Toomre parameter eq. \ref{eq:ToomreQ} the maximum growth rate can be rewritten as
\begin{equation}
    s_{\rm max}(R)
    =
    \kappa(R)
    \sqrt{
    \frac{1}{Q^2(R)} - 1
    }.
\end{equation}
Therefore, the perturbations grow fastest at the radius $R_\ast$ maximizing $s_{\rm max}(R)$, considering only the unstable regions where $Q(R)<1$.  

The characteristic size of the resulting fragments\footnote{It is important to distinguish this fastest-growing wavelength, $\lambda_{\rm max}$, from the
rotational Toomre scale, 
\begin{equation}
    \lambda_T(R)
    =
    \frac{2\pi}{k_T(R)}
    =
    \frac{\pi^2 G \Sigma_g(R)}{\Omega^2(R)}.
\end{equation}
which is derived from balancing self-gravity against rotational support.
Although $\lambda_T$ and $\lambda_{\rm max}$ are related, they are not
identical.} 
 is then estimated as
\begin{equation}
    \lambda_{\rm cl}(R_\ast)
    \simeq
    \lambda_{\rm max}(R_\ast)
    =
    \frac{2 \sigma^2}
    {G \Sigma_g(R_\ast)}.
\label{eq:Rcl}
\end{equation}

The corresponding clump mass
may be estimated as the gas mass contained
within a region of radius $\lambda_{\rm cl}/2$,
\begin{equation}
    M_{\rm cl}
    \sim
    \pi
    \left(
    \frac{\lambda_{\rm cl}}{2}
    \right)^2
    \Sigma_g(R_\ast),
\end{equation}
This yields
\begin{equation}
    M_{\rm cl}
    \sim
    \frac{\pi \sigma^4}
    {G^2 \Sigma_g(R_\ast)}.
\label{eq:Mcl}
\end{equation}

Combining eqs.~\ref{eq:Rcl} and \ref{eq:Mcl} yields a simple mass--size relation for the fragments produced by
the fastest-growing Toomre mode,
\begin{equation}
\boxed{M_{\rm cl}
=
\pi
\frac{\sigma^2}{G}
R_{\rm e},}
\label{eq:Mcl1}
\end{equation}
where we have introduced the effective clump radius,
$R_{\rm e} ={\lambda_{\rm cl}}/{2}$.
Thus, for a given gas velocity dispersion, the model predicts a linear relation between clump mass and size, i.e. $M_{\rm cl}\propto R_{\rm e}$.

Conversely, the observed masses and sizes of compact stellar clusters may be
used to infer the characteristic turbulent velocity dispersion of the parent
gas disk. Solving the above relation for $\sigma$ gives
\begin{equation}
\sigma
=
\left(
\frac{GM_{\rm cl}}
{\pi R_{\rm e}}
\right)^{1/2},
\end{equation}
or, in convenient astrophysical units,
\begin{equation}
\sigma
\simeq
37
\left(
\frac{M_{\rm cl}}
{10^6\,M_\odot}
\right)^{1/2}
\left(
\frac{R_{\rm e}}
{1\,{\rm pc}}
\right)^{-1/2}
{\rm km\,s^{-1}}.
\end{equation}

This value is remarkably similar to the turbulent gas velocity
dispersions, $\sigma \sim 30$--$80\,{\rm km\,s^{-1}}$, mostly inferred from
resolved [CII] and [OIII] kinematics of gas-rich galaxies at
$z\gtrsim4$ \citep[e.g.,][]{Ubler19,Rizzo21,Parlanti23,Roman23,Rowland24, Danhaive25}.
This agreement suggests that
the observed cluster properties may provide a direct probe of the turbulent
state of the gaseous disks from which they formed.
We caution, however, that the inferred dispersion is
tracer dependent. Using synthetic observations of galaxies at
$4\leq z\leq9$, \citet{Kohandel24} found that H$\alpha$ can yield
substantially larger velocity dispersions than [CII],
with $\sigma_{\rm H\alpha}>2\sigma_{\rm [CII]}$ in a significant
fraction of massive galaxies, because the two lines probe different
gas phases and kinematic components. For the present Toomre analysis, the [CII]-based dispersion is likely the more relevant quantity, since it more closely traces the kinematics of the thin gaseous disk that provides the turbulent support entering $Q$.

Observers often characterize compact stellar clusters by their effective mass
surface density. The expression for $\Sigma_{\rm cl}$ is obtained from its definition using again eqs.~\ref{eq:Rcl} and \ref{eq:Mcl}:  
\begin{equation}
\boxed{\Sigma_{\rm cl}
\equiv
\frac{M_{\rm cl}}
{\pi R_{\rm e}^2} =
\frac{\sigma^2}
{G R_{\rm e}},}
\label{eq:Sigcl}
\end{equation}
or
\begin{equation}
\Sigma_{\rm cl}
\simeq
5.8\times10^5
\left(
\frac{\sigma}
{50\,{\rm km\,s^{-1}}}
\right)^2
\left(
\frac{R_{\rm e}}
{1\, {\rm pc}}
\right)^{-1}
M_\odot\,{\rm pc^{-2}}.
\end{equation}

Thus, at fixed $\sigma$, the model predicts
\begin{equation}
\Sigma_{\rm cl}
\propto
R_{\rm e}^{-1},
\end{equation}
providing a second observable scaling relation linking the structural
properties of compact stellar clusters to the turbulent state of the
high-redshift gaseous disks from which they formed.

The mass--size relation eq. \ref{eq:Mcl1} does not by itself guarantee that a
clump with given $(M_{\rm cl},R_{\rm e})$ can actually form through Toomre
instability. Since gravitational fragmentation requires the parent disk to
satisfy the instability criterion $Q<1$, it is useful to express this
condition directly in terms of the observable clump properties.

Using eq.~(\ref{eq:Sigcl}), the Toomre parameter becomes
\begin{equation}
Q
=
\frac{\sigma\kappa}
{\pi G\Sigma_g}
=
\frac{\sigma\kappa R_{\rm e}^2}
{G M_{\rm cl}},
\end{equation}
where we have identified the gas surface density of the parent disk with the
initial surface density of the newly formed clump,
$\Sigma_g=\Sigma_{\rm cl}=M_{\rm cl}/(\pi R_{\rm e}^2)$.

Eliminating the turbulent velocity dispersion using the mass--size relation
eq.~\ref{eq:Mcl1} gives
\begin{equation}
Q
=
\frac{\kappa R_{\rm e}^{3/2}}
{\sqrt{\pi G M_{\rm cl}}},
\end{equation}
so that the Toomre instability condition becomes
\begin{equation}
M_{\rm cl}
>
\frac{\kappa^2}
{\pi G}
R_{\rm e}^3.
\label{eq:stable}
\end{equation}

\begin{center}
    \begin{figure*}
        \centering
    	\includegraphics[scale=0.57]{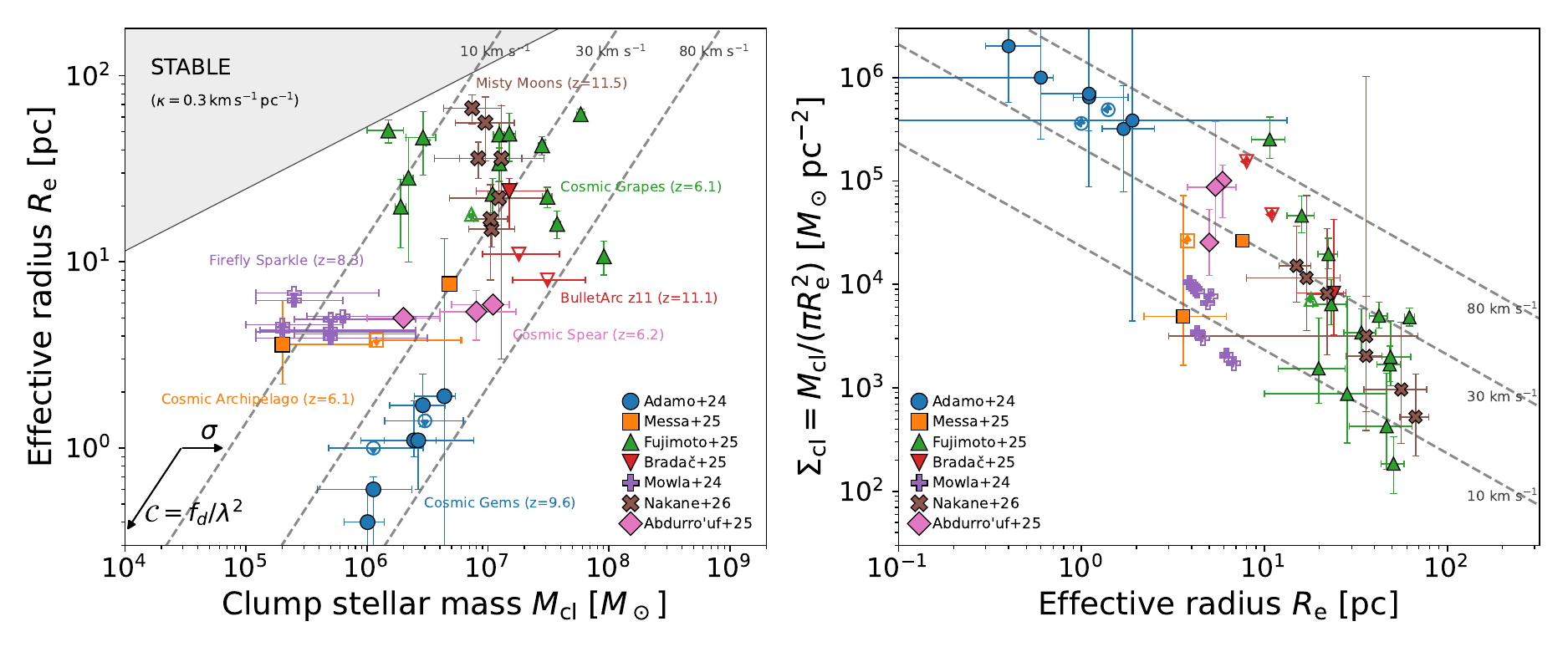}
    	\caption{ 
Observed mass--size and surface-density--size relations of compact stellar
clumps in high-redshift galaxies.
\textit{Left panel:} effective radius, $R_{\rm e}$, as a function of stellar
clump mass, $M_{\rm cl}$.
\textit{Right:} stellar mass surface density,
$\Sigma_{\rm cl}=M_{\rm cl}/(\pi R_{\rm e}^2)$, as a function of effective
radius.
Data points are from \citet[][]{Adamo24b, Messa25, Fujimoto25, Bradac25, Mowla24, Nakane25, Abdurrouf25} as shown in the legend. 
The dashed grey lines show the Toomre fragmentation model prediction from eqs. \ref{eq:Mcl1} and \ref{eq:Sigcl} for effective turbulent velocity dispersions
$\sigma=10$, $30$, and $80\,{\rm km\,s^{-1}}$. The observed clumps populate the region bracketed by
$\sigma\simeq10$--$80\,{\rm km\,s^{-1}}$, consistent with the turbulent
velocity dispersions inferred for gas-rich galaxies during the EoR. 
The grey area denotes the Toomre stable region where no clumps can form (see eq. \ref{eq:stable}), assuming a representative value $\kappa = 0.3\ \rm km\, s^{-1} pc^{-1}$. The schematic arrows illustrate the distinct roles of the model
parameters. At fixed $R_e$, increasing $\sigma$ shifts a system toward larger $M_{\rm cl}$, whereas increasing the disk
compactness $\mathcal{C}$ moves the fragment toward
smaller masses and radii along a constant-$\sigma$ sequence.  }
\label{fig:Mcl_vs_Re}
\end{figure*}
\end{center}

Eq.~(\ref{eq:stable}) defines a critical curve in the
$(M_{\rm cl},R_{\rm e})$ plane separating the region where Toomre
fragmentation is possible from the region where the parent disk remains
gravitationally stable. Clumps lying below this cubic relation would require
$Q>1$ and therefore cannot be produced by linear Toomre instability,
whereas clumps above the curve are compatible with fragmentation in
Toomre-unstable disks. In Fig.~\ref{fig:Mcl_vs_Re} we illustrate this
constraint by shading the Toomre-stable region for a representative value
$\kappa=0.3\,{\rm km\,s^{-1}\,pc^{-1}}$.

In comparing this relation with the observed stellar clumps, we
implicitly assume that the stellar system approximately preserves the
size of the initial Toomre fragment and that the integrated
star-formation efficiency is high, so that
$R_{e,\star}\simeq R_{\rm e}$ and $M_\star\simeq M_{\rm cl}$.
More generally, writing $M_\star=\epsilon_\star M_{\rm cl}$ and
$R_{e,\star}=f_R R_{\rm e}$ gives
\begin{equation}
M_\star =
\frac{\epsilon_\star}{f_R}
\frac{\pi\sigma^2}{G}R_{e,\star}.
\end{equation}
Thus a constant $\epsilon_\star/f_R$ changes only the normalization
of the mass--size relation, whereas mass-dependent star-formation
efficiency or structural evolution could modify its shape. We do not
model these later processes here.

It is important to emphasize that the mass--size relation
eq.~\ref{eq:Mcl1} does not predict a unique clump mass or size for a
given turbulent velocity dispersion. Rather, it defines a one-parameter
family of solutions corresponding to different gas surface densities in the
parent disk. 

Consequently, the dashed lines in Fig.~\ref{fig:Mcl_vs_Re} should not be
interpreted as predicting a unique clump mass for a given turbulent velocity
dispersion. Instead, they represent loci of constant $\sigma$ along which
the position of a given clump is determined by the surface density of the
parent gaseous disk. Low-surface-density disks fragment into fewer, more
extended and massive condensations, whereas high-surface-density disks
fragment on progressively smaller scales, producing more compact and less
massive clumps while remaining on the same constant-$\sigma$ sequence.

In the present model, the characteristic surface density of the gaseous disk
is primarily controlled by its compactness. We therefore introduce the
dimensionless compactness parameter
\begin{equation}
\mathcal{C}
\equiv
\frac{f_d}{\lambda^2}.
\label{eq:compac}
\end{equation}
Since
\begin{equation}
\Sigma_g
\propto
\frac{M_d}{R_d^2}
\propto
\frac{f_d\,M_{\rm vir}}
{\lambda^2R_{\rm vir}^2},
\end{equation}
it follows that, at fixed halo mass and redshift,
\begin{equation}
\Sigma_g \propto \mathcal{C}.
\end{equation}
Therefore, increasing the compactness parameter $\mathcal{C}$ shifts the
Toomre fragments toward the lower-left corner of the mass--size diagram,
yielding smaller and less massive clumps, while preserving the linear
relation $M_{\rm cl}\propto R_{\rm e}$ determined solely by the turbulent
velocity dispersion. Conversely, systems with lower compactness produce
larger and more massive Toomre fragments along the same constant-$\sigma$
sequence.

This interpretation, also supported by numerical simulations \citep[e.g.,][]{Mayer25}, naturally explains how compact stellar systems such as
the Cosmic Gems clusters can be reproduced with
$\sigma\simeq50$--$70\,{\rm km\,s^{-1}}$, provided that they form in
sufficiently compact, high-surface-density disks characterized by large
values of the compactness parameter $\mathcal{C}$.

\subsection{Jeans scale in the disk}
It is useful to define a local Jeans scale
associated with the fragmentation of individual Toomre condensations. Unlike the Toomre instability, which is regulated by the effective gas
velocity dispersion $\sigma$ (dominated by turbulence in high-redshift
galaxies), the Jeans scale is controlled by the thermal sound speed $c_s$.
The Jeans scale characterizes the smallest thermally supported condensations that can collapse within a Toomre fragment. It therefore governs the internal fragmentation of a clump rather than the initial fragmentation of the galactic disk.

The Jeans scale is defined as 
\begin{equation}
\lambda_J(R)
=
c_s t_{\rm ff}(R),
\end{equation}
where $t_{\rm ff}$ is the local free-fall time.
To compute $t_{\rm ff}$, we first estimate the midplane gas density from the
disk surface density and vertical scale height. Assuming a vertically
stratified disk,
\begin{equation}
\rho_{\rm mid}(R)
\simeq
\frac{\Sigma_g(R)}
{2H(R)},
\end{equation}
where $H(R)$ is the disk scale height. The free-fall time is
\begin{equation}
t_{\rm ff}(R)
=
\left(
\frac{3\pi}
{32G\rho_{\rm mid}(R)}
\right)^{1/2}.
\end{equation}

To estimate the scale height we assume vertical hydrostatic equilibrium.
Since the vertical support of gas-rich high-redshift disks is expected to be
dominated by turbulent rather than thermal pressure, the scale height is
taken to be
\begin{equation}
H(R)
\simeq
\frac{\sigma}
{\Omega(R)};
\end{equation}
the corresponding midplane density becomes
\begin{equation}
\rho_{\rm mid}(R)
\simeq
\frac{\Sigma_g(R)\,\Omega(R)}
{2\sigma}.
\end{equation}



Further using eq. \ref{eq:Sigcl} to eliminate $\Sigma_g$ yields
\begin{equation}
\lambda_J
=
c_s
\left(
\frac{3\pi}
{32}
\frac{\lambda_{\rm cl}}
{\sigma\Omega}
\right)^{1/2}.
\label{eq:lj}
\end{equation}

At fixed angular frequency and turbulent velocity dispersion,
the Jeans scale increases only as the square root of the Toomre scale,
so that progressively larger Toomre clumps contain larger Jeans fragments.
This hierarchy of
scales naturally leads to the expectation that each Toomre clump may
fragment into a population of smaller, thermally-supported condensations. The observational implications of this scenario are explored in \citet[][]{Behrendt16} and \citet[][]{Meng20}.
An upper limit to the number of thermally unstable sub-condensations is
\begin{equation}
N_{\rm frag}
\sim
\left(
\frac{\lambda_{\rm cl}}
{\lambda_J}
\right)^3.
\end{equation}
We point out that for typical parameters ($c_s = 1\ \kms, \sigma = 50\ \kms, \Omega = 0.3\ \kms \rm pc^{-1}$) eq. \ref{eq:lj} gives $\lambda_J = 0.08\ (\lambda_{\rm cl}/\rm pc)^{1/2}\ \rm pc$. 
{The adopted thermal sound speed $c_s=1\,{\rm km\,s^{-1}}$ is representative of warm ($T\simeq280$ K for molecular gas), metal-poor material; colder gas would yield proportionally smaller Jeans fragmentation scales.}

For large Toomre condensations, $\lambda_J$ can approach
parsec scales and may therefore become observable in strongly lensed
systems. However, the corresponding Jeans fragment \textit{masses} obtained
from the simple thermal estimate are much smaller than those of the
most massive compact clusters, such as the Cosmic Gems \citep[][]{Adamo24b, Vanzella26}. Thus, these
objects need not be interpreted as individual Jeans fragments; in the
present framework they can instead arise directly as Toomre fragments
of sufficiently compact disks. Hierarchical fragmentation followed by
accretion or coalescence remains a possible alternative.


\section{Clump mass spectrum}\label{sec:cl_spectrum}
{
Throughout this Sec., we distinguish the theoretical {Toomre fragment mass spectrum}, constructed from unstable disk modes, from the observed stellar-clump or star-cluster mass function. Their connection is not necessarily one-to-one, because the primary condensations may undergo secondary Jeans fragmentation (Sec.~2.3).
}
The fastest-growing Toomre mode provides a characteristic clump mass, but it does not describe the full distribution of masses expected to emerge from a gravitationally unstable disk. A more complete treatment can be obtained by considering the entire spectrum of unstable modes. 

\subsection{Intrinsic clump mass spectrum}
The number of unstable perturbations in an annulus of radius $R$ and width
$dR$ is proportional to the disk area,
$dA=2\pi R\,dR$. Integrating over radius and over the unstable wavenumber interval yields the
intrinsic {
Toomre fragment mass spectrum,
}
\begin{equation}
\begin{split}
\left(
\frac{dN}{d\log M}
\right)
&\propto
\int 2\pi R\,dR
\int d\ln k\, W(k,R) \\
&\times
\delta_D\!\left[
\log M-\log M(k,R)
\right],
\end{split}
\label{eq:spectrum}
\end{equation}
where $\delta_D$ is the Dirac delta function, and $W(k,R)$ is a weighting function measuring the efficiency with which a given unstable mode grows. {
We adopt a growth-rate weighting of the unstable modes,
\begin{equation}
W(k,R)\propto s(k,R),
\label{eq:weight_rate}
\end{equation}
where $s(k,R)$ is given by eq.~\ref{eq:growth_rate}. This choice assigns greater weight to modes with shorter linear instability timescales. We stress that linear theory determines the amplification rate of perturbations, not the number of nonlinear fragments: the mapping from unstable modes to fragment abundances remains phenomenological.
}

The resulting \textit{intrinsic} spectrum is shown as a dashed blue line in Fig. \ref{fig:spectrum} for the fiducial case $f_d=0.07$. It generally exhibits a broad peak around the mass corresponding to the fastest-growing mode, but possesses a finite width, typically extending for almost two dex, from $5.5 \simlt \log (M/M_\odot) \simlt 7.3$, that
reflects the entire range of unstable wavelengths. 

{
The resulting mode-weighted distribution is not a universal stellar-cluster mass function. Mass functions close to $dN/dM\propto M^{-2}$ are also found in nearby molecular clouds and stellar clusters; consequently, agreement with such a slope is not unique evidence for the present mechanism. The low-mass turnover of the primary Toomre spectrum should not be interpreted as a minimum mass for secondary Jeans fragments (Sec.~2.3).
}
Although the intrinsic mass spectrum is not strictly a power-law, its slope is not very different from the relation measured by \citet[][]{Claeyssens26}, $dN/dM\propto M^\beta$ with $\beta ={-1.89^{+0.13}_{-0.12}}$. We warn the reader that the comparison is only indicative as the \citet{Claeyssens26} sample combines clusters over a broad range of redshift and host-galaxy environments, while the spectrum in Fig. \ref{fig:spectrum} shows single-halo predictions with the parameters listed in the caption. Their sample is also restricted to young clusters with $M_{\rm cl} > 2\times 10^6\, M_\odot$, i.e. not a complete clump population over the same physical conditions.
In spite of these caveats, we show below that dynamical friction
steepens the high-mass tail of the predicted spectrum. {
The comparison with the slope measured by \citet{Claeyssens26} is an indicative consistency check, not a unique validation of the mode-weighting prescription.
}
Interestingly, the superposition of spectra from galaxies with
different compactness may also produce a population-averaged mass
function that is closer to a power law than the spectrum of any
individual galaxy.

\subsection{Correcting the spectrum for dynamical friction}
The intrinsic spectrum derived above describes the masses of clumps formed by
gravitational instability. However, the most massive clumps also experience
the strongest dynamical friction and therefore migrate most rapidly toward
the galactic center. Since such clumps are removed from the population of orbiting disk clumps, the observed mass spectrum differs from the formation spectrum. In the following we describe how we modify the intrinsic spectrum to account for dynamical friction effects.  

\subsubsection{Dynamical friction timescale}
A clump of mass $M_{\rm cl}$ moving through gas of density $\rho_g$ with velocity $v$
experiences a dynamical friction force
\begin{equation}
F_{\rm df}
\simeq
4\pi G^2 M_{\rm cl}^2 \rho_g
\frac{\ln\Lambda}{v^2}
\mathcal{I}(\mathcal{M}),
\end{equation}
where $\ln\Lambda$ is the Coulomb logarithm,
$\mathcal{M}= {v}/{c_s}$ is the Mach number, and $\mathcal{I}(\mathcal{M})$ is a dimensionless factor of order unity \citep[][see also \citealt{Pandey26}]{Ostriker99}. For supersonic motion, $\mathcal{I}\simeq 1$. The corresponding orbital decay timescale is
\begin{equation}
t_{\rm df}
\sim
\frac{M_{\rm cl} v}{F_{\rm df}}
\simeq
\frac{v^3}
{4\pi G^2 M_{\rm cl} \rho_g \ln\Lambda}.
\end{equation}
Since
$t_{\rm df}
\propto
M_{\rm cl}^{-1}$, massive clumps migrate inward much more rapidly than low-mass clumps.

\begin{center}
    \begin{figure}[!t]
        \centering
    	\includegraphics[scale=0.50]{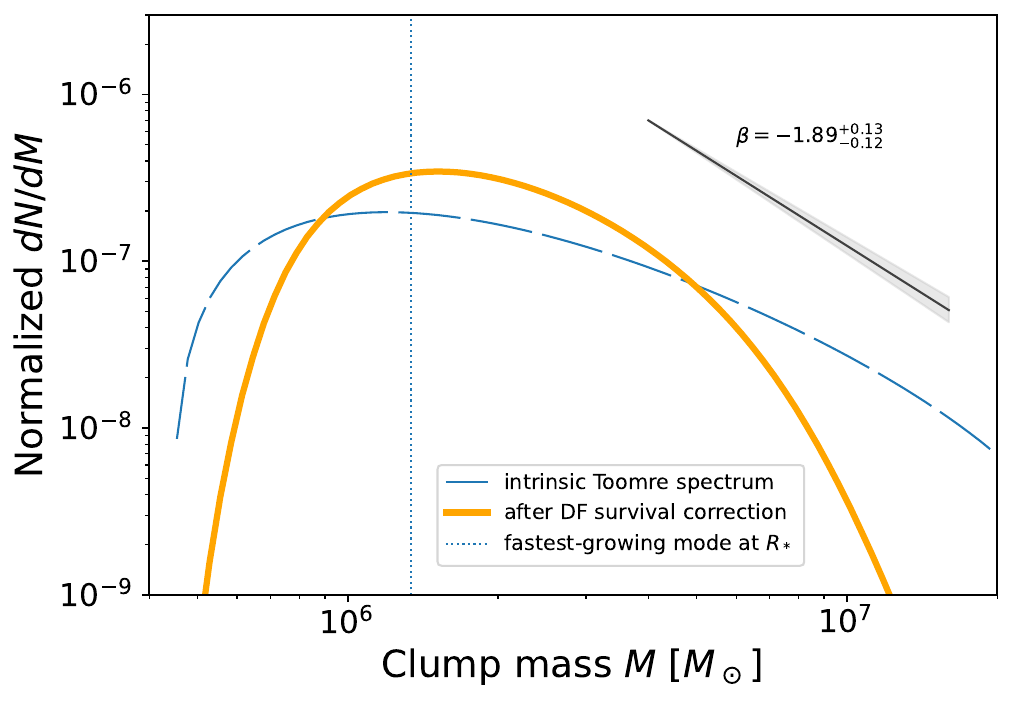}
    	\caption{
Normalized clump mass spectrum predicted by the fiducial model
($M_{\rm vir}=7\times10^{10}\,M_\odot$, $z=11$, $\lambda=0.035$,
$c=5$, $\sigma=10\,{\rm km\,s^{-1}}$, and $f_d=0.07$).
The blue dashed curve shows the {\it intrinsic} clump mass function obtained by integrating over the full spectrum of unstable Toomre modes.
The orange solid curve includes the additional effect of gas
dynamical friction, assuming an available migration time of
$t_{\rm avail}=70\,{\rm Myr}$.
The vertical dotted line marks the characteristic clump mass associated
with the fastest-growing Toomre mode at the radius of maximum growth,
$R_\ast$. The grey line in the upper-right corner indicates the
observed clump mass-function slope,
$dN/dM\propto M^\beta$ with $\beta \simeq {-1.89}$,
shown for reference only and with arbitrary normalization. 
}
        \label{fig:spectrum}
    \end{figure}
\end{center}

\begin{center}
    \begin{figure}[!t]
        \centering
    	\includegraphics[scale=0.50]{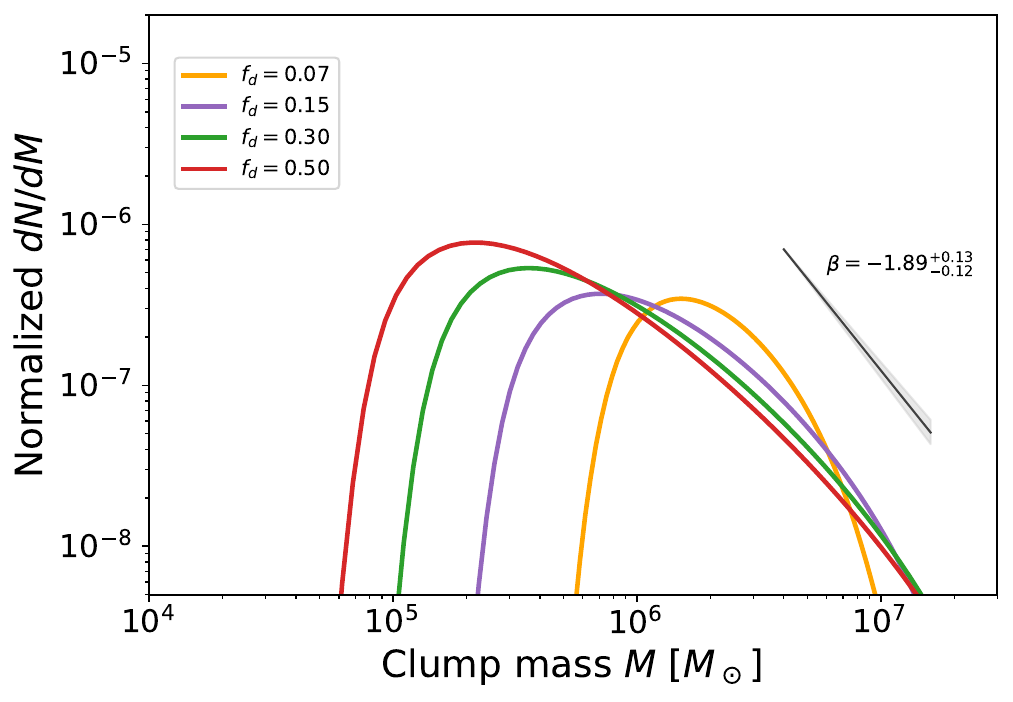}
    	\caption{
Same as Fig.~3, but showing the final clump mass spectrum after dynamical-friction correction for different
values of $f_d$ at fixed $\lambda=0.035$. Since
$\mathcal{C}=f_d/\lambda^2$, the sequence is equivalently a sequence of
increasing disk compactness. Increasing $f_d$ systematically shifts the mass distribution toward lower clump masses and increases its amplitude, reflecting the enhanced gravitational instability of more massive gas disks. Consequently, galaxies with larger disk mass fractions are expected to form a larger number of lower-mass clumps, whereas low-$f_d$ disks preferentially produce fewer, more massive condensations. 
}
    \label{fig:spectrum_fd}
    \end{figure}
\end{center}

{
For Figs.~\ref{fig:spectrum} and \ref{fig:spectrum_fd}, we set
$v(R)=V_c(R)$, the NFW halo circular velocity, and
$\rho_g(R)=\Sigma_g(R)/[2H(R)]$, with
$H(R)=\sigma/\Omega(R)$; we adopt $\ln\Lambda=3$. The migration mass satisfying
$t_{\rm df}[M_{\rm df}(R),R]=t_{\rm avail}$ is therefore
radius-dependent. For the fiducial model of Fig.~\ref{fig:spectrum},
at $R=R_d=0.276\,{\rm kpc}$ we obtain
$v=86.3\,{\rm km\,s^{-1}}$,
$\rho_g=9.29\,M_\odot\,{\rm pc^{-3}}$, and
$M_{\rm df}=1.39\times10^6\,M_\odot$ for
$t_{\rm avail}=70\,{\rm Myr}$. The mass spectra use the local
$t_{\rm df}(M,R)$ in the smooth survival probability defined below,
rather than a single mass cutoff.
}

\subsection{Critical migration mass}
Let $t_{\rm avail}$ denote the characteristic time over which the
clump population can evolve dynamically before substantial
replenishment or restructuring of the parent disk. This timescale
should not necessarily be identified with the interval between major
mergers alone. At high redshift, continuous cosmological accretion,
minor mergers, and disk reconfiguration may also replenish the gas
reservoir and reset the conditions for fragmentation. 

We therefore interpret $t_{\rm avail}$ as an effective coherence time of the
clump-forming disk. For the representative calculations below we adopt
$t_{\rm avail}=70\,{\rm Myr}$, comparable to the characteristic
$50$--$100\,{\rm Myr}$ evolutionary timescales of massive galaxies at
these redshifts.

A critical migration mass may be defined by the condition
\begin{equation}
t_{\rm df}(M_{\rm df})
=
t_{\rm avail}.
\end{equation}
Clumps with $M>M_{\rm df}$ are expected to reach the galactic center before the disk is replenished.

We stress that ``removal'' by dynamical friction does not imply
physical destruction of the clump. Rather, it denotes removal from
the population of orbiting disk clumps considered in the mass
function. Massive clumps reaching the central regions may survive as
bound systems and merge with other infalling clumps, contributing to
the build-up of a nuclear star cluster or compact bulge. Alternatively,
they may be partially or completely disrupted and deposit their stars
into the central diffuse component. Distinguishing between these
outcomes requires following the internal evolution and tidal response
of the clumps during inspiral, which is beyond the scope of the present
model.

\subsection{Survival probability}
Instead of imposing a sharp cutoff at $M_{\rm df}$, it is convenient to
introduce a smooth survival probability.
Following the general idea that cluster survival is a gradual rather than binary process (e.g. \citealt{Fellhauer07, Kruijssen15}), we model the survival probability by the smooth function
\begin{equation}
P_{\rm surv}(M,R)
=
\left[
1+
\left(
\frac{t_{\rm avail}}
{t_{\rm df}(M,R)}
\right)^\gamma
\right]^{-1},
\end{equation}
where $\gamma$ controls the sharpness of the transition. In the following, we fix $t_{\rm avail} = 70$ Myr and $\gamma=4$.
Hence, $P_{\rm surv}\rightarrow 1\, {\rm for}\, t_{\rm df} \gg t_{\rm avail}$, and
$P_{\rm surv} \rightarrow 0\, {\rm for} \,t_{\rm df}
\ll t_{\rm avail}.$

The observed clump mass function is finally obtained by replacing the weighting function in eq. \ref{eq:spectrum} with 
\begin{equation}
W(k,R)
\rightarrow
W(k,R)
P_{\rm surv}[M(k,R),R].
\end{equation}

Figure~\ref{fig:spectrum} illustrates the effect of gas dynamical
friction on the intrinsic Toomre mass spectrum for the fiducial model.
Because $t_{\rm df}\propto M_{\rm cl}^{-1}$, the most massive clumps migrate
toward the galactic centre much more rapidly than low-mass clumps.
Consequently, the high-mass tail of the intrinsic distribution is
preferentially suppressed\footnote{The precise mass scale at which dynamical friction suppresses the
spectrum depends on the adopted disk coherence time. Since
$t_{\rm df}\propto 1/M_{\rm cl}$, the corresponding migration mass
scales approximately as $M_{\rm df}\propto 1/t_{\rm avail}$; hence
even a factor of two uncertainty in $t_{\rm avail}$ shifts the
characteristic suppression scale by only $\simeq0.3$ dex without
altering the qualitative shape of the effect.}, while the low-mass end remains almost
unchanged. This selective removal shifts the peak of the surviving mass
function toward lower masses and steepens its high-mass decline,
bringing the predicted spectrum into remarkably good agreement with the
observed slope measured by \citet{Claeyssens26}.

Figure~\ref{fig:spectrum_fd} illustrates how the clump mass function depends
on the disk gas fraction, $f_d$, while keeping all other halo and disk
properties fixed.
As $f_d$ increases, the entire mass spectrum shifts
systematically toward lower clump masses,
indicating that more gas-rich disks produce fragments with
smaller characteristic masses. 

The observed clump mass function itself may not be universal.
While \citet{Claeyssens26} find
$\beta=-1.89^{+0.13}_{-0.12}$, the JADES sample \citep[][]{Zhu26} yields the somewhat
shallower value $\beta=-1.50^{+0.18}_{-0.17}$. Given the different
redshift and mass ranges, clump definitions, and selection functions,
a direct comparison is difficult. Such variations are qualitatively
expected in our framework: individual galaxies have curved rather
than pure power-law spectra, whose characteristic masses and
high-mass suppression depend on disk compactness, halo properties,
and the available dynamical-friction time. Consequently, the
effective power-law slope inferred over a finite observed mass range
need not be universal.


Figure~\ref{fig:spectrum_fd} illustrates the effect of increasing disk compactness, $\mathcal{C}=f_d/\lambda^2$, by varying $f_d$ at fixed
$\lambda=0.035$. Larger $\mathcal{C}$ shifts the clump mass spectrum
toward lower masses and increases the number of fragments.

This prediction is somewhat counterintuitive, as one might expect that disks
containing more gas would preferentially form more massive condensations.
Instead, this behaviour follows directly from the physics of Toomre fragmentation.
Increasing the disk mass fraction raises the gas surface density,
$\Sigma_g$, thereby reducing both the wavelength of the fastest-growing mode,
$\lambda_{\rm max}$, 
and the corresponding characteristic clump mass, $M_{\rm cl}$.
Thus, at fixed turbulent velocity dispersion, more massive gas disks become
unstable on progressively smaller spatial scales, naturally producing a
larger population of lower-mass clumps.

An additional implication of this picture concerns the temporal
variability of star formation. A gravitationally unstable disk does not
form a single stellar system, but a population of clumps spanning a
broad range of masses and growth times. Unless all unstable modes reach
the non-linear regime simultaneously, star formation will therefore
proceed through a sequence of spatially distinct episodes, naturally
providing a physical route to \textit{bursty} galaxy-wide star formation (see e.g. \citealt{Pallottini23}).
This interpretation is consistent with the cloud-scale model of
\citet{Cueto26}, in which the discrete formation of star-forming
clouds naturally generates stochastic and bursty star-formation
histories \citep[][]{Simmonds25, Bevins26}, and simulations \citet{Horie26}. 

Since increasing disk compactness produces a larger number of
lower-mass fragments, it may also reduce the fractional amplitude of
galaxy-wide SFR fluctuations by averaging over a larger number of
independent star-forming regions. Conversely, less compact disks
forming fewer clumps may exhibit more strongly bursty integrated
star-formation histories, providing a potentially testable connection
between galaxy compactness, clump multiplicity, and burstiness.

Recent JADES results \citep[][]{Zhu26} also suggest that the stellar clump mass--size
relation may depend on galactocentric radius, with a somewhat
shallower relation for inner than for outer clumps. Such a trend could
reflect radial variations in the initial Toomre fragmentation
conditions, but may also encode subsequent structural evolution during
inward migration. Our present dynamical-friction treatment modifies
the abundance of orbiting clumps but does not follow changes in their
mass or radius, preventing a quantitative prediction of this effect.
Resolved radial statistics therefore provide an interesting future
test of the model.

\section{Feedback and clump visibility}
\label{sec:feedback}
\begingroup
The instability discussed in Sec.~\ref{sec:formation} first produces gaseous
Toomre fragments. Whether their stellar components emerge as observable compact
clumps depends on the evolution and possible dispersal of the residual gas and
dust. Here we revisit the momentum budget provided by different mechanical and radiative feedback processes, namely supernova explosions, radiation pressure on dust and Ly$\alpha$ radiation pressure. 

The force comparisons below are simple {one-zone tests for global gas removal}. They neglect several possible complications such as disruption of individual dense filaments, directional
outflows through low-density channels, or the subsequent expansion of the
stellar component. Hence, they represent a necessary local acceleration
criterion. This approach has precedents in the literature,
{including \citet{Dekel13}; our focus is its application to extremely compact Toomre fragments.}

For a consistent comparison of the feedback channels, we adopt an
instantaneous stellar burst at $Z=0.001$, using STARBURST99
\citep[SB99,][]{Leitherer99} spectra, ionizing photon rates, and mechanical
energy histories, and assuming a Salpeter IMF
spanning $0.1$--$100\,M_\odot$. We evaluate the feedback over $1$--$40$ Myr after the burst.

\subsection{Mass accounting and gravitational force}
\label{sec:feedback_gravity}
We denote by \(M_{\rm cl}\) the initial gas mass of a Toomre fragment, as derived in Sec. 2. A fraction \(\epsilon_\star\) of this mass is converted into stars, such that \(M_\star=\epsilon_\star M_{\rm cl}\), leaving a residual gas mass \(M_g=(1-\epsilon_\star)M_{\rm cl}\). In the preceding sections, we have implicitly identified the Toomre fragment mass with the stellar clump mass, corresponding to the high-efficiency limit \(\epsilon_\star\simeq1\). Here we relax this approximation to determine whether stellar feedback permits such efficient conversion. For simplicity, we retain the characteristic clump radius \(R_e\) throughout the embedded phase.

The gravitational force
opposing removal of the residual gas is approximated by
\begin{equation}
 F_{\rm grav}\simeq\frac{G M_g(M_g+M_\star)}{R_{\rm e}^2}
 =\frac{G(1-\epsilon_\star)M_{\rm cl}^2}{R_{\rm e}^2}.
 \label{eq:feedback_force_correct}
\end{equation}
Using the Toomre relation (eq.~\ref{eq:Mcl1}) we find
\begin{equation}
 F_{\rm grav}\simeq(1-\epsilon_\star)\frac{\pi^2\sigma^4}{G}.
 \label{eq:feedback_force_toomre}
\end{equation}
At fixed $R_{\rm e}$, the surface density of the residual gas is
\begin{equation}
 \Sigma_g=(1-\epsilon_\star)\Sigma_{\rm cl},\qquad
 \Sigma_{\rm cl}=\frac{\sigma^2}{G R_{\rm e}}.
 \label{eq:feedback_surface_gas}
\end{equation}

{It is important to distinguish the integrated star-formation efficiency of an individual clump, $\epsilon_\star$, from the galaxy-wide efficiency of converting halo baryons into stars. If a fraction $f_d$ of the halo baryons settles into the disk, and a fraction $f_{\rm cl}$ of this disk gas participates in star-forming clumps, the efficiency associated with a single fragmentation episode is
\begin{equation}
    \epsilon_{\rm gal}
    \equiv
    \frac{M_{\star}^{\rm gal}}{f_b M_{\rm vir}}
    \simeq
    f_d\,f_{\rm cl}\,\epsilon_\star .
\end{equation}
Even in the limiting case $\epsilon_\star=1$, representative values $f_d\simeq0.15$ and $f_{\rm cl}\simeq0.5$ yield $\epsilon_{\rm gal}\simeq0.075$. Similarly, the simulations of \citet{Horie26}, with $f_d=0.065$ and $f_{\rm cl}\simeq0.2$, imply $\epsilon_{\rm gal}\lesssim0.013$. These estimates illustrate how efficient star formation within individual clumps can coexist with a modest instantaneous conversion of the available halo baryons into stars. However, gas expelled from clumps may remain bound to the galaxy and participate in subsequent star-formation episodes. The cumulative galaxy-wide efficiency can therefore be higher and depends on gas recycling, accretion, and the duration of star formation, which are not modeled here.}

\subsection{Gas dispersal by supernova explosions}
\label{sec:SN_dispersal}
The supernova momentum injection rate can be expressed as
\begin{equation}
 \dot p_{\rm SN}(t)=p_{\rm SN}\dot N_{\rm SN}(t),
 \label{eq:sn_momentum_new}
\end{equation}
where the terminal momentum of an isolated remnant is approximated by
$p_{\rm SN}=1.2\times10^5\,M_\odot\,{\rm km\,s^{-1}}$ at the dense-gas
conditions discussed in \citet[][]{Thornton98,Martizzi15}, and the time-dependent SN rate is computed self-consistently from SB99. The corresponding force ratio is
\begin{equation}
 \Gamma_{\rm SN}(t)=\frac{p_{\rm SN}\dot N_{\rm SN}(t)}{F_{\rm grav}}.
 \label{eq:sn_burst_gamma}
\end{equation}
For $\sigma=50\,{\rm km\,s^{-1}}$, $R_{\rm e}=1$ pc, and
$\epsilon_\star=0.8$, we find
$\Gamma_{\rm SN,max}\simeq0.019$ at $t\simeq3.9$ Myr,
indicating that SNe cannot globally expel the residual gas in this case.

\subsection{Gas dispersal by radiation pressure on dust}
\label{sec:RP_dust}
The direct dust-mediated momentum input is $\dot p_{\rm dust}=f_{\rm abs}L_{\rm bol}/c$,
with the fraction of stellar light absorbed by dust written as
{
\begin{equation}
 f_{\rm abs}=1-e^{-\tau_{\rm UV}},\qquad
 \tau_{\rm UV}=\kappa_{\rm UV}\,{D}\,\Sigma_g,
 \label{eq:dust_uv_abs}
\end{equation}
where $D$ is the dust-to-gas mass ratio in Milky Way units and
$\kappa_{\rm UV}=778\,{\rm cm^2\,g^{-1}}$ is the UV
opacity.}
For comparison, we consider also an idealized case of maximal trapping of IR photons, which may substantially
overestimate the actual momentum transferred to the gas. In this case,
{
\begin{equation}
 \dot p_{\rm dust}^{\rm IR}=(f_{\rm abs}+\tau_{\rm IR})\frac{L_{\rm bol}}c,
 \label{eq:dust_ir_benchmark}
\end{equation}
where the IR optical depth is calculated in Appendix~\ref{ap:tauIR}.
Multidimensional radiation-hydrodynamic results
\citep[][]{Menon22,Menon24} indicate that $\tau_{\rm IR}$ is not generally
realized as an effective momentum boost. Nevertheless, we retain the IR-enhanced case as a useful {benchmark}.}

An instantaneous stellar population has a time-dependent bolometric luminosity,
$L_{\rm bol}(t)=\ell_{\rm bol}(t)M_\star$. We determine $\ell_{\rm bol}(t)$
by integrating the SB99 stellar spectrum over wavelength at each
burst age. The resulting force ratios are
\begin{equation}
 \begin{aligned}
 \Gamma_{\rm dust}^{\rm single}(t)&=
 \frac{f_{\rm abs}L_{\rm bol}(t)}{cF_{\rm grav}},\\
 \Gamma_{\rm dust}^{\rm IR}(t)&=
 \frac{(f_{\rm abs}+\tau_{\rm IR})L_{\rm bol}(t)}{cF_{\rm grav}}.
 \end{aligned}
 \label{eq:dust_burst_gamma}
\end{equation}
Importantly, $\tau_{\rm UV}$ and $\tau_{\rm IR}$ are computed from the
\emph{remaining} gas column $\Sigma_g$, rather than from the initial one.
For $D=0.01$, $\epsilon_\star=0.8$, $\sigma=50\,{\rm km\,s^{-1}}$,
and $R_{\rm e}=1$ pc, the maximum direct and maximal-IR force ratios
are $\Gamma_{\rm dust,max}^{\rm single}\simeq0.0086$ and
$\Gamma_{\rm dust,max}^{\rm IR}\simeq0.019$, respectively. Both are
sub-Eddington under these conditions.

\subsection{Gas dispersal by Ly$\alpha$ radiation pressure}
\label{sec:feedback_lya}
Resonantly scattered Ly$\alpha$ photons can impart a larger momentum than
direct single scattering, according to
\begin{equation}
 \dot p_\alpha(t)={\cal M}_F\frac{L_\alpha(t)}c.
 \label{eq:lya_burst_force}
\end{equation}
We adopt the force-multiplier prescription 
based on \citet{Tomaselli21} and its numerical radiative-transfer tests
\citep[][]{Nebrin26}:
\begin{equation}
 \begin{split}
 {\cal M}_F&=\min\left[10^{-5}
 \left(\frac{N_{\rm HI}}{{\rm cm}^{-2}}\right)^{1/3},\, 35.2\,(T_4D)^{-1/4}\right],
 \end{split}
 \label{eq:MF_minimum}
\end{equation}
where the dust-limited term is omitted for $D=0$ and
\begin{equation}
 N_{\rm HI}=\frac{M_g}{\pi R_{\rm e}^2\mu_{\rm H}m_p}
 =\frac{(1-\epsilon_\star)\sigma^2}
 {G R_{\rm e}\mu_{\rm H}m_p}.
 \label{eq:MF}
\end{equation}
We adopt $\mu_{\rm H}=1.4$ and assume the residual gas is neutral. The column-density contribution to $\mathcal M_F$ decreases as
$\epsilon_\star$ approaches unity. 
Large multipliers are supported in idealized neutral, optically thick
systems; their dynamical efficacy in dusty, porous clouds
remains uncertain \citep{Menon26}.

The ionizing photon rate, $Q_0(t)$, of the burst is again obtained from 
SB99. Assuming complete absorption of ionizing photons
and Case-B recombination,
\begin{equation}
 L_\alpha(t)=0.68\,h\nu_\alpha\,Q_0(t),
 \label{eq:lya_sb99_lum}
\end{equation}
where $h\nu_\alpha=10.2\,{\rm eV}$. The time-dependent force ratio is
\begin{equation}
 \Gamma_\alpha(t)=\frac{{\cal M}_F L_\alpha(t)}{c F_{\rm grav}}.
 \label{eq:lya_sb99_gamma}
\end{equation}
With $M_g$, $R_{\rm e}$, and $N_{\rm HI}$ held fixed after star formation,
$\Gamma_\alpha(t)$ evolves through $Q_0(t)$.

\begin{figure*}
 \centering
 \includegraphics[width=\textwidth]{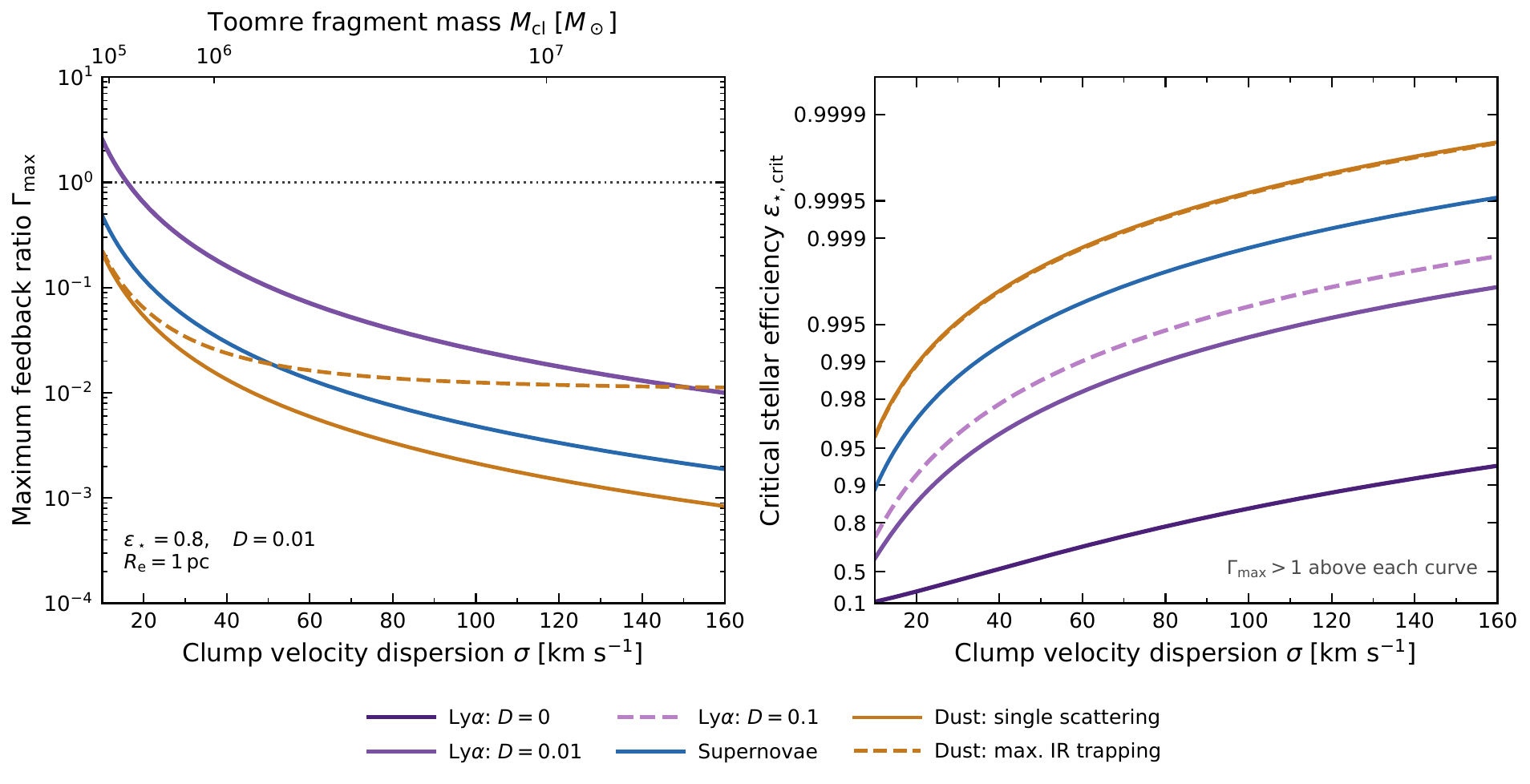}
\caption{Feedback in Toomre fragments with $R_{\rm e}=1$ pc.
\textit{Left:} Maximum Eddington ratio
$\Gamma_{\max}=\max_t[\dot p(t)/F_{\rm grav}]$ over $1$--$40$ Myr
as a function of $\sigma$, for $\epsilon_\star=0.8$ and $D=0.01$.
The purple, blue, and solid orange curves show Ly$\alpha$, supernovae,
and single-scattering dust radiation pressure, respectively; the dashed
orange curve includes maximal IR trapping. The dotted horizontal line
marks $\Gamma_{\max}=1$, and the upper axis gives $M_{\rm cl}$.
\textit{Right:} Critical stellar efficiency $\epsilon_{\star,\rm crit}$
at which $\Gamma_{\max}=1$ versus $\sigma$. The dark, medium, and
light purple Ly$\alpha$ curves correspond to $D=0$, $0.01$, and $0.1$,
respectively (the last is dashed). The blue curve shows supernovae;
the solid and dashed orange curves show dust single scattering and
maximal IR trapping, respectively, for $D=0.01$.
The vertical scale is logarithmic in $1-\epsilon_{\star,\rm crit}$;
the super-Eddington regime lies above each curve.}
\label{fig:feedback}\end{figure*}

\subsection{Feedback comparison}
\label{sec:feedback_comparison}

Figure~\ref{fig:feedback} compares the maximum force ratios achieved
by each feedback mechanism during the evolution of an instantaneous
burst. The left panel shows $\Gamma_{\max}$ as a function of the
clump velocity dispersion at $R_{\rm e}=1$ pc,
$\epsilon_\star=0.8$, and $D=0.01$; the upper axis translates
$\sigma$ into the initial Toomre fragment mass through
eq.~\ref{eq:Mcl1}. At $\sigma=10\,{\rm km\,s^{-1}}$,
Ly$\alpha$ becomes super-Eddington ($\Gamma_{\alpha,\max}\simeq2.56$),
while SNe and dust radiation pressure remain sub-Eddington. At
$\sigma=50\,{\rm km\,s^{-1}}$, corresponding to
$M_{\rm cl}\simeq1.83\times10^6\,M_\odot$,
$\Gamma_{\alpha,\max}\simeq0.102$,
$\Gamma_{\rm SN,max}\simeq0.019$, and the direct and maximal-IR
dust benchmarks give $\Gamma_{\max}\simeq0.0086$ and $0.019$,
respectively. Thus, the more massive fragments are increasingly
resistant to global gas removal.

The right panel shows the critical integrated stellar efficiency
$\epsilon_{\star,\rm crit}$ for which $\Gamma_{\max}=1$. The
Ly$\alpha$ curves explore different dust abundances, whereas the SN
and dust curves use $D=0.01$. For $\sigma=50\,{\rm km\,s^{-1}}$,
Ly$\alpha$ becomes super-Eddington at
$\epsilon_{\star,\rm crit}\simeq0.616$, $0.975$, and $0.986$
for $D=0$, $0.01$, and $0.1$, respectively; the corresponding
thresholds for SNe and direct dust pressure are $0.995$ and $0.998$
at $D=0.01$. At this dust abundance, the Ly$\alpha$ threshold
rises from $\epsilon_{\star,\rm crit}\simeq0.862$ at
$\sigma=20\,{\rm km\,s^{-1}}$ to $\simeq0.994$ at
$\sigma=100\,{\rm km\,s^{-1}}$. Hence, substantial gas conversion
is generally required before feedback can overcome gravity in the
more massive fragments.

We conclude that compact Toomre fragments are very resistant to
gas removal by stellar feedback under representative
conditions. Ly$\alpha$ radiation pressure offers the most favorable
route to super-Eddington acceleration, particularly in dust-poor
gas or at high $\epsilon_\star$. As a caveat, we note that sub-Eddington
conditions also do not exclude directional outflows or the removal
of diffuse, low-column-density gas, whose evolution requires
radiation-hydrodynamic modeling.

The increasing resistance of massive Toomre fragments to
gas removal provides an \emph{a posteriori} consistency check
of the approximation $M_\star\simeq M_{\rm cl}$ adopted in
Sec.~\ref{sec:formation}. In particular, the weakness of
stellar feedback suggests that most of the initial fragment
mass can be converted into stars, favoring very high
star-formation efficiencies.
\endgroup

\section{Summary} \label{sec:summary}
The unprecedented angular resolution of JWST has revealed that galaxies
during the EoR are composed of compact star-forming
clumps spanning a broad range of masses and sizes. While the Toomre
instability has long been recognized as the leading mechanism for
fragmentation of gas-rich disks, most previous studies have focused on
the characteristic mass associated with the fastest-growing mode rather
than on the full clump population expected to emerge from gravitational
instability.

In this work we have developed a self-consistent analytical framework
linking the global properties of high-redshift galactic disks to the
formation, dynamical evolution and visibility of stellar clumps.
Starting from the classical Toomre instability, we derived both the
scaling relations and the intrinsic mass spectrum of unstable
fragments. We then followed the modification of this population by gas
dynamical friction and finally investigated whether stellar feedback
can disperse the natal gas and dust, allowing the resulting stellar
systems to become observable.
Our main results are:
\begin{itemize}
\item[\color{red}$\blacksquare$]
{ Toomre fragmentation is favored in gas-rich, compact disks
with moderate turbulence. The instability threshold depends
sensitively on the velocity dispersion and disk gas fraction,
making clump formation a natural, but not universal,
outcome in super-early galaxies.}

\item[\color{red}$\blacksquare$] The fastest-growing Toomre mode predicts the scaling relations
$M_{\rm cl}\propto R_{\rm e}$ and
$\Sigma_{\rm cl}\propto R_{\rm e}^{-1}$. The observed clumps are
consistent with turbulent velocity dispersions
$\sigma\simeq10$--$80\,{\rm km\,s^{-1}}$. At fixed $\sigma$, the location
of a fragment along these relations is determined by the disk
compactness,
$\mathcal{C}=f_d/\lambda^2$.

\item[\color{red}$\blacksquare$] The intrinsic Toomre mass spectrum extends over nearly two orders of
magnitude,
$5.5\lesssim\log(M_{\rm cl}/M_\odot)\lesssim7.3$,
demonstrating that gravitational instability naturally produces a broad
distribution of fragment masses rather than a single characteristic
scale.

\item[\color{red}$\blacksquare$] Gas dynamical friction preferentially drives the most massive clumps toward the galactic centre
($t_{\rm df}\propto M_{\rm cl}^{-1}$), suppressing the high-mass tail
of the orbiting clump population and {
yielding a high-mass tail broadly consistent with, but not uniquely constrained by, the observed stellar-clump mass-function slope,
}
$\beta=-1.89^{+0.13}_{-0.12}$.

\item[\color{red}$\blacksquare$] Increasing the disk compactness, $\mathcal{C}=f_d/\lambda^2$, shifts the characteristic Toomre
mass toward lower values while increasing the number of fragments.
Compact galaxies are therefore predicted to host larger populations of
smaller stellar clumps, providing a direct observational test of the
model.

\item[\color{red}$\blacksquare$]
{
Once formed, massive Toomre fragments are remarkably resistant
to gas removal by supernovae, dust and Ly$\alpha$ radiation pressure.
Feedback typically becomes super-Eddington only after
$\gtrsim 95\%$ of the initial gas has been converted into stars,
suggesting very high star-formation efficiencies and supporting
the approximation $M_\star\simeq M_{\rm cl}$.
}\end{itemize}
The present work follows the formation of clumps by Toomre instability,
the subsequent modification of their population by gas dynamical
friction, and the conditions under which their stellar populations might become observable
{despite substantial resistance of the natal gas to global feedback-driven expulsion}. We deliberately do not follow the later dynamical response of
the stellar component to gas removal, including expansion, tidal
stripping, and long-term dissolution, as these processes require
coupling the orbital evolution of the clusters to their internal
structure. These effects, together with comparisons to dedicated
radiation-hydrodynamic simulations currently in preparation, will be
presented in forthcoming work.

\section*{Acknowledgments}
We thank M. Nakane for sharing key data and comments; A. Adamo, O. Agertz, Y. Asada, M. Brada{\v{c}}, S. Fujimoto, V. Kokorev, M. Messa, Y. Nakazato, A. Romeo, S. Tacchella, E. Vanzella, Y. Zhu for useful discussions and comments. We are grateful to the organizers and the participants of the Nordita 2026 Workshop `\textit{Zoom-in Views of Galaxy Disks Across Cosmic Time}' -- where part of this work has been carried out -- for stimulating interactions. AF acknowledges support from the ERC Advanced Grant INTERSTELLAR H2020/740120. Partial support from the Carl Friedrich von Siemens-Forschungspreis der Alexander von Humboldt-Stiftung Research Award is kindly acknowledged (AF). This research was partly supported (AF) by grant NSF PHY-2309135 to the Kavli Institute for Theoretical Physics (KITP) and the Munich Institute for Astro-, Particle and BioPhysics (MIAPbP) which is funded by the Deutsche Forschungsgemeinschaft under Germany´s Excellence Strategy – EXC-2094 – 390783311. The authors acknowledge the use of ChatGPT (OpenAI) during manuscript preparation, and language editing. The authors take full responsibility for all scientific content, interpretations, and conclusions. We gratefully acknowledge computational resources of the Center for High Performance Computing (CHPC) at SNS.

\bibliographystyle{apsrev4-1}

\bibliography{arxiv}

\begin{appendix}\label{app:IRtrapping}
\section{A. Infrared trapping in dusty Toomre clumps}
\label{ap:tauIR}
\begingroup

The effect of reprocessed radiation on a dusty clump depends on the IR
optical depth of the {residual gas}. For the idealized case of maximal IR trapping,
the effective momentum-coupling factor is
{
\begin{equation}
 f_{\rm trap}^{\rm IR}=f_{\rm abs}+\tau_{\rm IR},
 \label{eq:ftrap_IR}
\end{equation}
where $f_{\rm abs}=1-\exp(-\tau_{\rm UV})$.}
The infrared depth is
\begin{equation}
 \tau_{\rm IR}=\kappa_{\rm IR}(T_d,D)\Sigma_g,
 \qquad \Sigma_g=(1-\epsilon_\star)\Sigma_{\rm cl},
 \label{eq:tauIR_def}
\end{equation}
where $D$ is normalized to the Milky Way dust-to-gas ratio ($D_{\rm MW}=162$), and
\begin{equation}
 \kappa_{\rm IR}\simeq5D\left(\frac{T_d}{100\,{\rm K}}\right)^2
 \;{\rm cm^2\,g^{-1}}.
 \label{eq:kappaIR}
\end{equation}

Using the initial Toomre-fragment relation
$\Sigma_{\rm cl}=\sigma^2/(G R_{\rm e})$, we find
\begin{equation}
 \Sigma_{\rm cl}\simeq 1.21\times10^2
 \left(\frac{\sigma}{50\,{\rm km\,s^{-1}}}\right)^2
 \left(\frac{R_{\rm e}}{1\,{\rm pc}}\right)^{-1}
 \;{\rm g\,cm^{-2}}
 \label{eq:SigmaIR_cgs}
\end{equation}
and hence
\begin{equation}
 \begin{aligned}
 \tau_{\rm IR}&\simeq6.1\times10^2
 (1-\epsilon_\star)D
 \left(\frac{T_d}{100\,{\rm K}}\right)^2
 \left(\frac{\sigma}{50\,{\rm km\,s^{-1}}}\right)^2
 \left(\frac{R_{\rm e}}{1\,{\rm pc}}\right)^{-1}.
 \end{aligned}
 \label{eq:tauIR}
\end{equation}

For dust absorption at stellar UV wavelengths,
\begin{equation}
 \tau_{\rm UV}=\kappa_{\rm UV}{D}
 (1-\epsilon_\star)\Sigma_{\rm cl},\\
 \label{eq:dust_uv_appendix}
\end{equation}
where the UV opacity is $\kappa_{\rm UV}=1.26\times10^5/D_{\rm MW} = 778\;{\rm cm^2\,g^{-1}}$.
The corresponding post-burst force ratio is
\begin{equation}
 \Gamma_{\rm dust}^{\rm IR}(t)=
 \frac{\ell_{\rm bol}(t)\epsilon_\star M_{\rm cl}}
 {c\,G(1-\epsilon_\star)M_{\rm cl}^2/R_{\rm e}^2}
 \left[1-e^{-\tau_{\rm UV}}+\tau_{\rm IR}\right].
 \label{eq:GammaDust}
\end{equation}

In the regime $\tau_{\rm UV}\gg1$, the factor in square brackets
reduces to $1+\tau_{\rm IR}$. We emphasize that this is a
\emph{maximal-coupling benchmark}, not a rigorously calibrated
upper limit. Turbulent inhomogeneity, anisotropic leakage and
dust-temperature variations can substantially reduce the coupling
compared with a simple $\tau_{\rm IR}L_{\rm bol}/c$ enhancement
\citep[][]{Menon22,Menon24}. In Fig.~\ref{fig:feedback}, the
time-dependent $\ell_{\rm bol}(t)$ is obtained by integrating the
STARBURST99 instantaneous-burst spectrum over wavelength at each age.
The residual gas mass and clump radius, and hence the optical depths,
are held fixed during the post-burst evolution.
\endgroup
\end{appendix}

\end{document}